\documentclass[twocolumn,showpacs,amsmath,amssymb]{revtex4}

\usepackage{graphicx}
\usepackage{dcolumn}
\usepackage{bm,color}

\begin{document}

\preprint{APS/}

\title{Neutrino potential for neutrinoless double beta decay}

\author{Yoritaka Iwata$^{1,2}$}
 \email{iwata_phys@08.alumni.u-tokyo.ac.jp}

\affiliation{$^{1}$Institute of Innovative Research, Tokyo Institute of Technology, Japan}
\affiliation{$^{2}$Department of Mathematics, Shibaura Institute of Technology, Japan}
\date{\today}

\begin{abstract}
Neutrino potential for neutrinoless double beta decay is studied with focusing on its statistical property.
The statistics provide a gross view of understanding amplitude of constitutional components of the nuclear matrix element.
\end{abstract}


\maketitle

\section{Introduction}
Observation of neutrinoless double beta decay is associated with important physics; e.g., 
\begin{itemize}
\item existence of Majorana particle,  
\item breaking of leptonic number conservation.  
\end{itemize}
In this sense neutrinoless double-beta decay is intriguing enough to bring about an example exhibiting the physics beyond the standard model of elementary particle physics. 
 Among several topics as for the double beta decay, it plays a role in 
\begin{itemize}
\item quantitative determination of neutrino mass,
\end{itemize}
where it is worth noting that neutrino is treated as massless particle in the standard model.

There is a relation between the half life of neutrinoless double-beta decay and the effective neutrino mass ($m_{\nu}$): 
\begin{equation} \begin{array}{ll}
　　[T_{0 \nu}^{1/2}] ^{-1} = G |M^{0 \nu}|^2\left (\frac{m_{\nu}}{m_e} \right)^2,　 
\end{array}  \end{equation}
where $G$ is the phase space factor (its value is obtained rather precisely), $m_e$ is the electron mass (its value is also precisely obtained), and $M^{0 \nu}$ is the nuclear matrix element (NME, for short). 
In order to determine the neutrino mass, it is necessary to calculate $M^{0 \nu}$ very precisely. 
Since the detail information on initial and final states (i.e., quantum level structure of these states) is necessary for the calculation of NMEs, it is impossible to have reliable NME without knowing nuclear structures. 
The impact of precise NME calculations is expected to be large enough (e.g., for a large-scale shell model calculation, see Ref.~\cite{15iwata}), and the unknown leptonic mass-hierarchy and the Majorana nature of neutrinos are expected to be discovered. 

As seen in the following the neutrino potential appears in the calculation of NMEs.
In this paper neutrino potential for neutrinoless double beta decay is studied from a statistical point of view.

\section{Neutrino potential}
\subsection{Nuclear matrix element}
Nuclear matrix element in double beta decay is investigated under the closure approximation.
It approximates all the different virtual intermediate energies by a single intermediate energy (i.e., with the averaged energy called closure parameter).
For neutrinoless double beta decay, nuclear matrix element is written by 
\begin{equation} \label{matrixel} \begin{array}{ll}
M^{0 \nu} = M_{\rm F}^{0 \nu} - \frac{g_V^2}{g_A^2} M_{\rm GT}^{0 \nu} + M_{\rm T}^{0 \nu}
\end{array} \end{equation}
where $g_V$ and $g_A$ denote vector and axial coupling constants, and $\alpha$ of $M_{\alpha}^{0 \nu}$ is the index for the double beta decay of three kinds: $\alpha =$ F, GT, T (Fermi, Gamow-Teller, and tensor parts). 
According to Ref.~\cite{iwata-cns},  each part is further represented by the sum of two-body transition density (TBTD) and anti-symmetrized two-body matrix elements. 
\begin{equation} \label{me-general} \begin{array}{ll}
M_{\alpha}^{0 \nu}  = \langle 0_f^+ |O_{\alpha}^{0 \nu} | 0_i^+ \rangle \vspace{2.5mm} \\
= \sum {\rm TBTD}(n'_1 l'_1 j'_1 t'_1, n'_2 l'_2 j'_2 t'_2, n_1 l_1 j_1 t_1,  n_2 l_2 j_2 t_2; J)  \vspace{1.5mm} \\
\quad \langle n'_1 l'_1 j'_1 t'_1, n'_2 l'_2 j'_2 t'_2; J|O_{\alpha}^{0 \nu}(r) | n_1 l_1 j_1 t_1, n_2 l_2 j_2 t_2; J  \rangle_{\rm AS}
\end{array} \end{equation}
where $O^{0 \nu}_{\alpha}(r)$ are transition operators of neutrinoless double beta decay, and $0_i^+$ and $0_f^+$ denote initial and final states, respectively.
The sum is taken over indices $( n_i l_i j_i t_i,n'_j l'_j j'_j t'_j)$ with ($i,j=1,2$), where $n$, $l$, $j$ and $t$ mean principal, angular momentum and isospin quantum numbers, respectively, $j_1$ and $j_2$ (or $j_1'$ and $j_2'$) are coupled to $J$ (or $J$), similarly $l_1$ and $l_2$ (or $l_1'$ and $l_2'$) are coupled to $\lambda$ (or $\lambda'$), and $t_1 = t_2 = 1/2, ~ t_1' = t_2' = -1/2$ is valid if neutrons decay into protons.

\begin{figure*} 
\includegraphics[width=180mm]{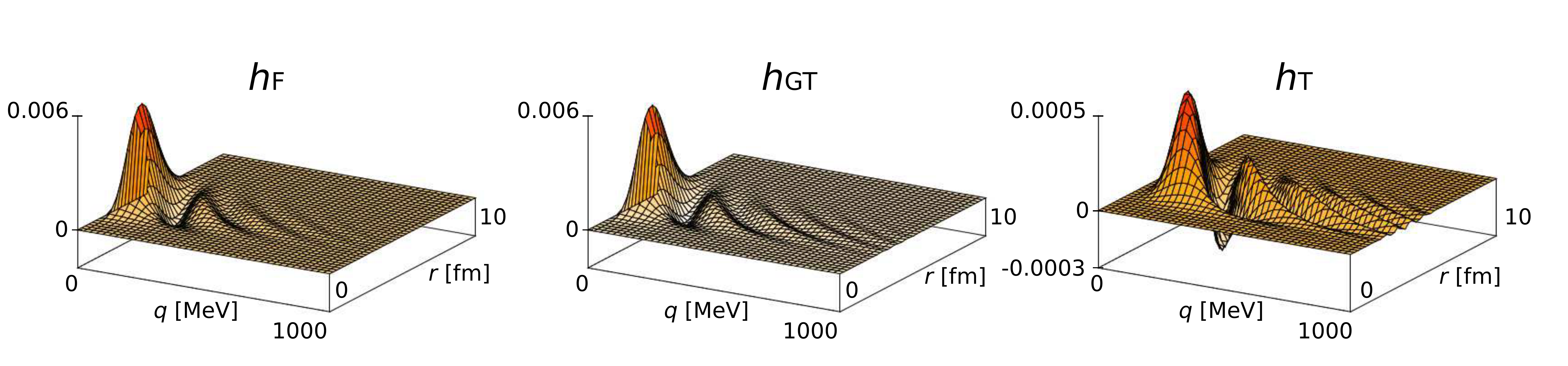}  \\
(a) $n=n' =0$ and $l=l'=3$ \\
\includegraphics[width=180mm]{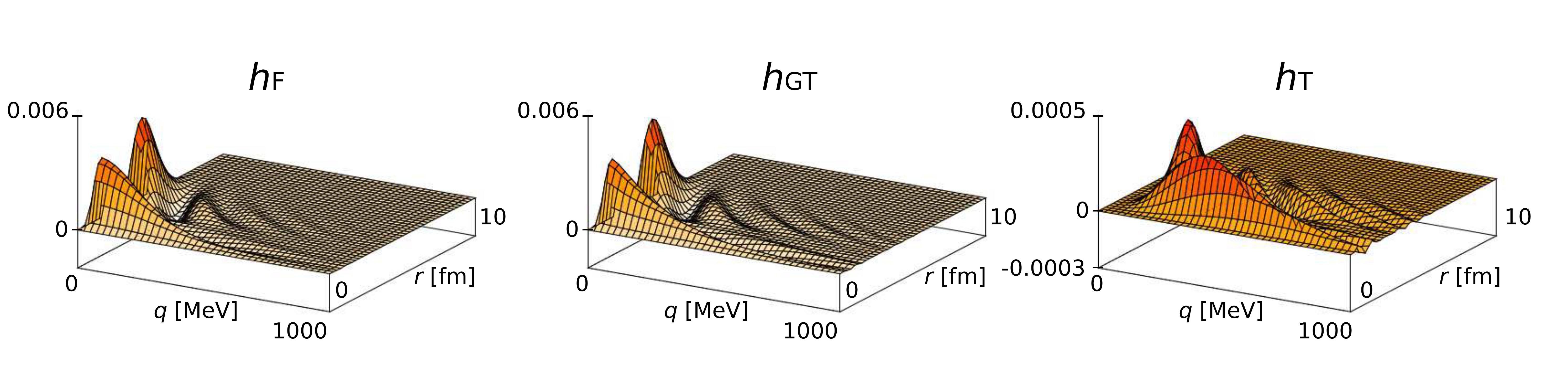}  \\
(b) $n=n' =1$ and $l=l'=0$ \\
\caption{ \label{fig1} (Color online)
Integrands of Eq.~\eqref{nupot2} are depicted for $n=n' =0$ and $l=l'=3$ in panel a, and for $n=n' =1$ and $l=l'=0$ in panel b.
The plots are made for $r=\sqrt{2}\rho=0$ to 10~fm and $q=0$ to 1000~MeV.
The closure parameter $\langle E \rangle$ is fixed to 0.5~MeV, which is suggested by the calculation without using closure approximation~\cite{13senkov}.
}
\end{figure*}

The two-body matrix element before the anti-symmetrization is represented by
\begin{equation} \begin{array}{ll}
\langle n'_1 l'_1 j'_1 t'_1, n'_2 l'_2 j'_2 t'_2; J|O_{\alpha}^{0 \nu}(r) | n_1 l_1 j_1 t_1, n_2 l_2 j_2 t_2; J  \rangle  \vspace{2.5mm} \\
 = 2\sum_{S, S', \lambda, \lambda'} \sqrt{j_1' j_2' S' \lambda'} \sqrt{j_1 j_2 S \lambda}  \vspace{1.5mm} \\
\quad \langle l_1' l_2' \lambda' S'; J| S_{\alpha} |  l_1 l_2 \lambda S; J \rangle 
~ \langle n_1' l_1' n_2' l_2'; J| H_{\alpha}(r) |  n_1 l_1 n_2 l_2 \rangle  \vspace{1.5mm} \\
\quad \left\{
\begin{array} {ccc}
l_1' & 1/2 & j_1' \\
l_2' & 1/2 & j_2' \\
\lambda' & S' & J 
\end{array}
\right\} 
~ \left\{
\begin{array} {ccc}
l_1 & 1/2 & j_1 \\
l_2 & 1/2 & j_2 \\
\lambda & S & J 
\end{array}
\right\}
\end{array} 
\end{equation}
where $H_{\alpha}(r)$ is the neutrino potential,  $S_{\alpha}$ denotes spin operators, $S$ and $S'$ mean the two-body spins, and $\{ \cdot \}$ including nine numbers denotes the 9j-symbol.
By implementing the Talmi-Moshinsky transforms:
\begin{equation}
 \langle n l, NL|  n_1 l_1, n_2 l_2 \rangle_{\lambda} 
 \langle n' l', N'L'|  n_1' l_1', n_2' l_2' \rangle_{\lambda'}  
\end{equation}
 the harmonic oscillator basis is transformed to the center-of-mass system.
\begin{equation} \begin{array}{ll}
 \langle l_1' l_2' \lambda' S'; J| S_{\alpha} |  l_1 l_2 \lambda S; J \rangle 
 \langle n_1' l_1' n_2' l_2'; J| H_{\alpha}(r) |  n_1 l_1 n_2 l_2 \rangle \vspace{3.5mm} \\
= {\displaystyle \sum_{n, n', l, l',N,N'}} \langle n l, NL|  n_1 l_1, n_2 l_2 \rangle_{\lambda} 
 \langle n' l', N'L'|  n_1' l_1', n_2' l_2' \rangle_{\lambda'}   \vspace{1.5mm} \\
 \langle l' L \lambda' S'; J|  S_{\alpha} | l L \lambda S; J \rangle 
 \langle n' l'|  H_{\alpha}(\sqrt{2} \rho) | n l \rangle, 
\end{array}  \end{equation}
where $\rho = r /\sqrt{2}$ is the transformed coordinate of center-of-mass system.
In this paper we focus on the neutrino potential effect arising from
\begin{equation} \label{nupot} \begin{array}{ll}
 \langle n' l'|  H_{\alpha}(\sqrt{2} \rho) | n l \rangle.
\end{array}  \end{equation}
This part is responsible for the amplitude of each transition from a state with $n$, $l$ to another state with $n'$, $l'$, while the cancellation is determined by spin-dependent part.

\begin{figure*} 
\includegraphics[width=185mm]{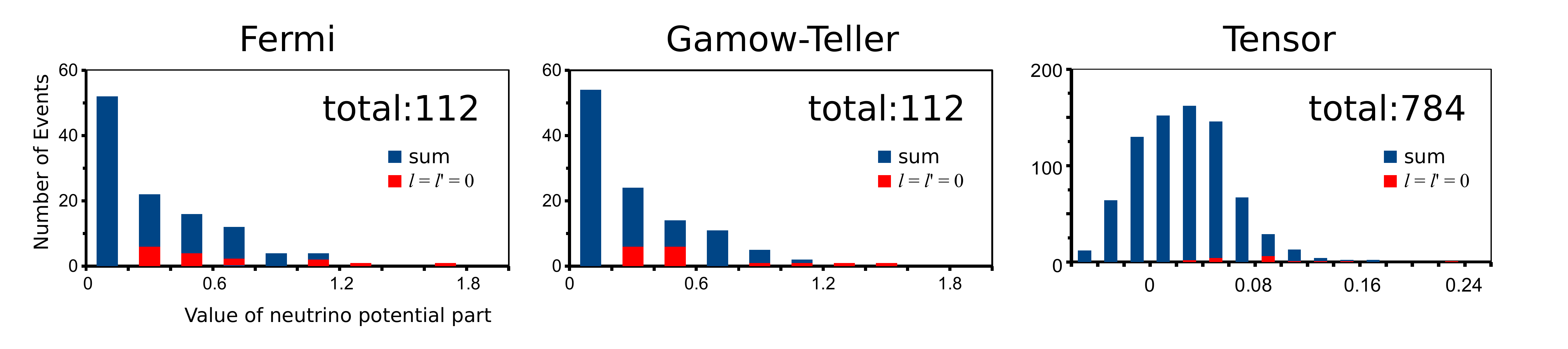} 
\caption{ \label{fig2} (Color online)
Frequency distribution of $\langle n' l'|  H_{\alpha}(\sqrt{2} \rho) | n l \rangle$ is shown limited to nonzero cases.
Cases with $n, n' = 0, 1, \cdots, 3$ and  $l, l' = 0, 1, \cdots, 6$ are taken into account, where note that $l\ne l'$ results in $\langle n' l'|  H_{\alpha}(\sqrt{2} \rho) | n l \rangle = 0$ in Fermi and Gamow-Teller cases~\cite{iwata-cns}.
The total number of events with nonzero $\langle n' l'|  H_{\alpha}(\sqrt{2} \rho) | n l \rangle$ is shown in each panel.
}
\end{figure*}

\begin{table*}[t]
\caption{Large contributions are listed from 1st to 10th largest ones. 
Two symmetric cases resulting in an equivalent value are shown in the same position for the tensor part with $l \ne l'$. }
\begin{center}
  \begin{tabular*}{\columnwidth}{@{\extracolsep{\fill}}c | cc| c c| c c} \hline \hline
 &  \multicolumn{2}{c|}{Fermi} &  \multicolumn{2}{c|}{Gamow-Teller} &  \multicolumn{2}{c}{Tensor} \\ 
Ranking &  $(n~ l~ n'~ l')$  & Value &  $(n~ l~ n'~ l')$  & Value & $(n~ l~ n'~ l')$ & Value  \\ \hline
 1   & (0~0~0~0) &  $1.626$ &  (0~0~0~0)  &  $1.488$ &  (0~0~0~0) & 0.2249  \\
 2  &  (1~0~1~0) & $1.307$ &  (1~0~1~0) &  $1.227$ &  (0~0~0~1) &  $0.1637$  \\
  &  &  &   &   &  (0~1~0~0) &  \\
 3   & (2~0~2~0) & $1.133$  &  (2~0~2~0)  & $1.081$ &  (1~0~1~0) & $0.1579$  \\ 
 4   & (0~1~0~1) & $1.126$  &  (0~1~0~1)  & $1.051$ &  (0~1~0~1) & $0.1435$  \\
 5   & (3~0~3~0) & $1.018$  &  (3~0~3~0)  & $0.982$ &  (2~0~2~0) & $0.1248$  \\  
 6   & (1~1~1~1) & $1.006$  &  (1~1~1~1)  & $0.937$ &  (0~0~1~1) & $0.1204$  \\  
   &  &  &   &   &  (1~1~0~0) &    \\
  7   & (2~1~2~1) & $0.922$  &  (2~1~2~1)  & $0.861$ &  (1~1~1~1) & $0.1203$  \\  
  8   & (0~2~0~2) & $0.899$  &  (0~2~0~2)  & $0.859$ &  (0~1~0~2) & $0.1130$  \\  
   &  &  &   &   &  (0~2~0~1) &    \\
  9   & (3~1~3~1) & $0.859$  &  (3~1~3~1)  & $0.805$ &  (1~0~1~1) & $0.1115$  \\
   &  &  &   &   &  (1~1~1~0) &   \\    
 10   & (1~2~1~2) & $0.836$  &  (1~2~1~2)  & $0.790$ &  (0~0~0~2) & $0.1112$  \\
   &  &  &   &   &  (0~2~0~0) &    \\    
\hline  \hline
 \end{tabular*}
\end{center}
\label{table1}
\end{table*}

\subsection{Neutrino potential represented in the center-of-mass system}
We pay special attention to the neutrino potential part~\eqref{nupot}.
Under the closure approximation neutrino potential at the massless neutrino limit~\cite{91tomoda,10horoi, 13senkov} is
\begin{equation} \label{nupot2} \begin{array}{ll}
H_{\alpha}(\sqrt{2} \rho) = \frac{2R}{\pi} \int_0^{\infty} f_{\alpha} (\sqrt{2} \rho q) \frac{h_{\alpha(q)}}{q+ \langle E \rangle} ~q ~ dq, 
\end{array} \end{equation}
where $q$ is the momentum of virtual neutrino, $R$ denotes the radius of decaying nucleus, and $f_{\alpha}$ is a spherical Bessel function ($\alpha=0,2$), 
In particular $\langle E \rangle$ is called the closure parameter, which means the averaged excitation energy of virtual intermediate state.
For the usage of ordinary light neutrinos, the neutrino potential in the massless limit is sufficient.
 In Eq.~\eqref{nupot2} neutrino potentials include the dipole form factors (not just the form factors) that take into account the nucleon size.
The representation of neutrino potentials are
\begin{equation} \label{eq-form} \begin{array}{ll}
h_{\rm F}(q^2) = \frac{g_V^2}{(1+q^2/\Lambda_V^2)^4}  \vspace{2.5mm} \\
h_{\rm GT}(q^2) =  \frac{2}{3} \frac{q^2}{4 m_p^2} (\mu_p - \mu_n) ^2 \frac{g_V^2}{(1+q^2/\Lambda_V^2)^4}  \\
\quad +
 \left( 1-\frac{2}{3} \frac{q^2}{q^2+m_{\pi}^2} + \frac{1}{3} \left( \frac{q^2}{q^2+m_{\pi}^2} \right)^2 \right)
 \frac{g_A^2}{(1+q^2/\Lambda_A^2)^4} 
 \vspace{2.5mm} \\
h_{\rm T}(q^2) =  \frac{1}{3} \frac{q^2}{4 m_p^2} (\mu_p - \mu_n) ^2 \frac{g_V^2}{(1+q^2/\Lambda_V^2)^4}  \\
\quad +
\left( \frac{2}{3} \frac{q^2}{q^2+m_{\pi}^2} - \frac{1}{3} \left( \frac{q^2}{q^2+m_{\pi}^2} \right)^2 \right) 
 \frac{g_A^2}{(1+q^2/\Lambda_A^2)^4} 
 \vspace{2.5mm} \\
\end{array} \end{equation}
where $\mu_p$ and $\mu_n$ are magnetic moments satisfying $\mu_p - \mu_n = 4.7$, $m_p$ and $m_{\pi}$ are proton mass and pion mass, and $\Lambda_V=850$MeV, $\Lambda_A=1086$MeV are the finite size parameters.

\begin{figure*} 
\includegraphics[width=180mm]{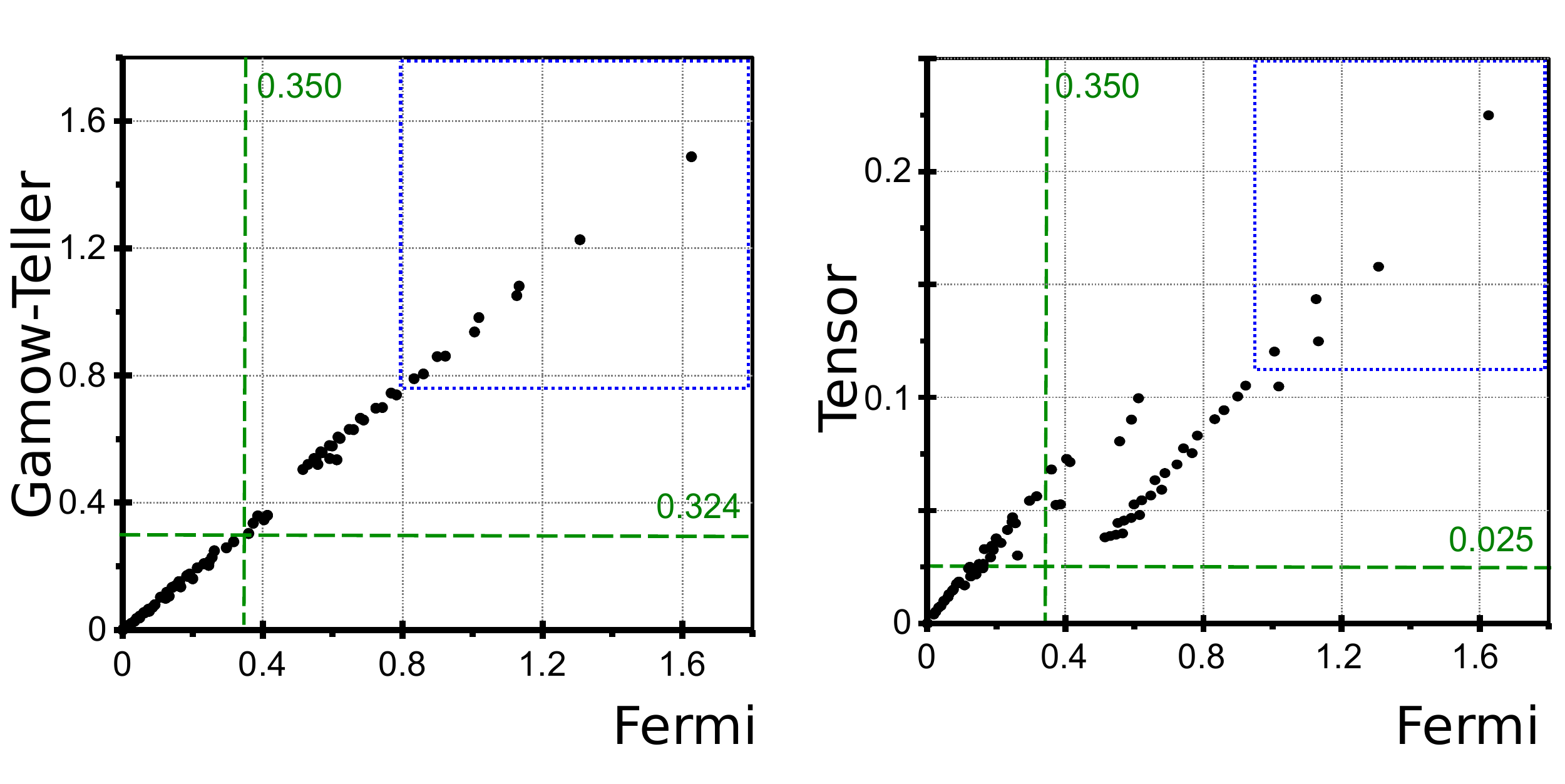} 
\caption{ \label{fig3} (Color online)
Correlation between Eq.~\eqref{nupot2} values are examined by assuming $l=l'$.
[Left] Correlation between Eq.~\eqref{nupot2} values for Fermi and Gamow-Teller parts, where the condition $l=l'$ does not bring about any limitations for Fermi and Gamow-Teller parts. 
[Right] Correlation between Eq.~\eqref{nupot2} values for Fermi and tensor parts, where values for the tensor part is always positive if $l=l'$ is assumed. 
For both panels, top 10 contributions listed in Table~\ref{table1} are included in dotted-blue rectangles, and the average of all the nonzero contributions are shown in green dashed lines.
}
\end{figure*}

Figure~\ref{fig1} shows the integrand of Eq.~\eqref{nupot2}.
In any case ripples of the form: $q \rho$ = const. can be found if $q$ and $\rho$ are relatively large.
The upper-value of the integral range should be at least equal to or larger than $q=1000$.  
In our research including our recent publication~\cite{15iwata}, we take $q=2000$~MeV and $r=10$~fm as the maximum value for numerical integration of Eq.~\eqref{nupot2} (massless neutrino cases).
This setting: $q_{\rm max} =2000$~MeV  and $r_{\rm max}=10$~fm is sufficient to obtain 3-digit accuracy of the nuclear matrix element.

\section{Statistics}
Since actual quantum states are represented by the superposition of basic states such as $| nl \rangle$ in the shell-model treatment, the contribution of neutrino potential part can be regarded as the superposition:
\begin{equation} \begin{array}{ll}
 {\displaystyle \sum_{n, n', l, l'}}  k_{n, n', l, l'} ~ \langle n' l'|  H_{\alpha}(\sqrt{2} \rho) | n l \rangle.
\end{array} \end{equation}
using a suitable set of coefficients $\{ k_{n, n', l, l'} \}$ determined by the nuclear structure of grandmother and daughter nuclei.
Accordingly it is worth investigating the statistical property of neutrino potential part~\eqref{nupot}. 

Frequency distribution of neutrino potential part~\eqref{nupot} is shown in Fig.~\ref{fig2}.
The values are always positive for Fermi and Gamow-Teller parts, while the tensor part includes non-negligible negative values.
Indeed, the sum of positive and negative contributions of tensor part suggests that total sum 19.88 is obtained by the cancellation between $+23.128$ and $-3.248$ (i.e., $19.880 = 23.128-3.248$).
The order of the magnitude is different only for the tensor part.
Indeed, the average of the nonzero components is 0.350 for the Fermi part, 0.324 for the Gamow-Teller part, and 0.025 for the tensor part.
Contributions with $l= l' =0$ (sum) cover 27.1$\%$ of the total contributions (sum) for Fermi and Gamow-Teller parts, and 7.2$\%$ for the tensor part.

Large contributions for Fermi, Gamow-Teller and tensor parts are summarized in Table~\ref{table1}.
Contribution labeled by $(n~ l~ n'~ l')=(0~0~0~0)$ (i.e. transition between $0s$ orbits) provides the largest contribution in any part. Roughly speaking, we see that $s$-orbit ($l=0$ or $l'=0$) plays a significant role.
The order of the kind $(n~ l~ n'~ l')$ are exactly the same for Fermi and Gamow-Teller parts as far as the top 10 list is concerned.
Ten largest contributions (sum) cover 45.0$\%$ of the total contributions (sum) for the Fermi part, 46.1$\%$ for the Gamow-Teller part, and 10.1$\%$ for the tensor part. 
The minimum value for the tensor part is -0.0450 achieved by $(n~ l~ n'~ l')=(0~4~1~0)$ and $(1~0~0~4)$.

Correlation between the values of Eq.~\eqref{nupot2} for different parts are examined in Fig.~\ref{fig3}. 
Comparison between Fermi and Gamow-Teller parts shows that they provide almost the same values, although the Fermi part generally shows slightly larger value compared to the Gamow-Teller part.
Such an quantitative similarity between Fermi and Gamow-Teller parts is not trivial since we can find essentially different mathematical representations at least in their form factors (cf.~Eq.~\ref{eq-form}). 
The tensor part is positively correlated with the Fermi part (therefore Gamow-Teller part).
The $l=l'$ components of the tensor part contributions (sum) cover 26.0$\%$ of the total tensor part contributions (sum).

\section{Summary}
There are components of the two kinds in the nuclear matrix element; one is responsible for the amplitude and the other is for the cancellation.
As a component responsible for the amplitude, neutrino potential part (i.e., Eq.~\eqref{nupot}) is investigated in this paper.
The presented results are valid not only to a specific double-beta decay candidates but also to all the possible candidates within $n, n' = 0, 1, \cdots, 3$ and  $l, l' = 0, 1, \cdots, 6$.
Note that, in terms of the magnitude, almost 40$\%$ smaller values are applied for the Gamow-Teller part in calculating the nuclear matrix element since $(g_V/g_A)^2 = (1/1.27)^2 \sim 0.62$ (cf. Eq.~\eqref{matrixel}).

Among several results, positive correlation of the values between Fermi, Gamow-Teller and tensor parts has been clarified.
Apart from the tensor part values, almost a half of the total contributions has been shown to be occupied only by 10 largest contributions, and 27$\%$ of the total contribution has been found out to be occupied by the $l =l'=0$ contributions.

The other components of the NMEs also responsible for the cancellation will be studied in the next opportunity.  \\

The author would like to express his sincere gratitude to Dr. J. Men\'endez for fruitful comments.
Numerical calculations were carried out at the workstation system of Institute of Innovative Research, Tokyo Institute of Technology, and COMA of University of Tsukuba.

\end{document}